\documentclass[preprint,aps,amsmath,pre]{revtex4-1}
\usepackage{graphicx}
\newcommand{\dps}{\displaystyle}

\begin{document}

\title{Robustness Against Extinction by Stochastic Sex Determination in 
Small Populations}

\author{David M. Schneider$^1$, Eduardo do Carmo$^1$, Yaneer
Bar-Yam$^2$ and Marcus A.M. de Aguiar$^{1,2}$}

\affiliation{$^1$ Instituto de F\'{\i}sica `Gleb Wataghin',
Universidade Estadual de Campinas, Unicamp\\ 13083-859, Campinas, SP,
Brasil\\$^2$New England Complex Systems Institute, Cambridge,
Massachusetts 02142}

\begin{abstract}

Sexually reproducing populations with small number of individuals may
go extinct by stochastic fluctuations in sex determination, causing all
their members to become male or female in a generation. In this work we
calculate the time to extinction of isolated populations with fixed
number $N$ of individuals that are updated according to the Moran birth
and death process. At each time step, one individual is randomly
selected and replaced by its offspring resulting from mating with
another individual of opposite sex; the offspring can be male or female
with equal probability. A set of $N$ time steps is called a
generation, the average time it takes for the entire population to be
replaced. The number $k$ of females fluctuates in time, similarly to a
random walk, and extinction, which is the only asymptotic possibility,
occurs when $k=0$ or $k=N$. We show that it takes only one
generation for an arbitrary initial distribution of males and females to
approach the binomial distribution. This distribution, however, is
unstable and the population eventually goes extinct in $2^N/N$
generations. We also discuss the robustness of these results against
bias in the determination of the sex of the offspring, a characteristic
promoted by infection by the bacteria Wolbachia in some arthropod species
or by temperature in reptiles.

\end{abstract}

\maketitle

%%%%%%%%%%%%%%%%%%%%%%%%%%%%%%%%%%%%%%%%%%%%%%%%%%%%%%%%%
%%%%%%%%%%%%%%%%%%%%%%%%%%%%%%%%%%%%%%%%%%%%%%%%%%%%%%%%%
\section{Introduction}

Most species in our planet have small numbers of individuals
\cite{rosen95}. Even the human population, now with more than seven
billion people, has gone through periods of very low abundances not too
long ago \cite{huff10}. In fact,typical abundance distributions of
several groups of species has been shown that follow a universal lognormal
curve with an excess of rare species \cite{rosen95,hubb01}.

Small communities are prone to extinction for a large number of
reasons, such as inability to protect themselves or difficulty in
finding mates \cite{ste99}, accumulation of deleterious mutations
\cite{lyn95} and shear stochastic fluctuations in the environment
\cite{drake09,men10,ova10} or in the number of males and females in the
group \cite{mel08}. If male and female offspring are equally likely to
occur, the ratio between males and females fluctuates over generations
\cite{eng07} and it might occur that they all become male or female,
driving the population to extinction. This works against the
persistence of small communities and it is somewhat puzzling how so
many such populations do exist. In some species, the females developed
ways to control the sex ratio of their progeny, producing more females in
a male rich environment and vice-verse. This mechanism, termed local mate
competition \cite{ham67,cha82}, may have evolved to avoids extinction by
fluctuations in sex determination.

In this work we consider the changes in the sex ratio of a population with
fixed number of individuals in which a single member is randomly selected
to reproduce at each time step, being replaced by a male or a female
offspring with equal probability. The model is a great simplification of
the dynamics of real populations, but it captures the random character of
the process and sheds light on why low abundance species are actually much
more robust than one could naively expect. In particular, we do not take
into account the fact that females usually have many offspring and, even
if the population size is held fixed, there is competition among the
offspring and it is the best fit individual who survives. In a population
consisting mostly of males, it is expected that a female offspring will
fare better than a male, contributing to balance the sex ratio. We shall
not take competition, natural selection, spatial structures \cite{rei12}
or aging into account \cite{mel08}, restricting our work to {\it neutral}
evolution, which has been shown to describe the observed universal
patterns of abundance and diversity, both for hermaphroditic and
sexual populations \cite{agu09,agu11,bap12}.

%%%%%%%%%%%%%%%%%%%%%%%%%%%%%%%%%%%%%%%%%%%%%%%%%%%%%%%%%
%%%%%%%%%%%%%%%%%%%%%%%%%%%%%%%%%%%%%%%%%%%%%%%%%%%%%%%%%
\section{The dynamical system}

We consider a population with $N$ individuals divided into $k$ females
and $N-k$ males. We call $P_t(k)$  be the probability of finding $k$
females at time $t$. The population is updated at discrete time steps
similarly to the birth and death process proposed by Moran in
population genetics \cite{moran}: at each time step a random individual
is selected to reproduce with an available member of the opposite sex;
after reproduction the selected individual dies and is replaced by the
offspring, which can be male or female with equal probability. The
states $P(0)$ and $P(N)$ are absorbing states corresponding to {\it all
males} and {\it all females} respectively and mark the extinction of
the population, since reproduction becomes impossible when all members
have the same sex. In a state with $k$ females, the probability of
having $k-1$ females in a single time step is $\Omega_{k,k-1}=k/2N$,
since one of the females has to be selected (probability $k/N$) and be
replaced by a male offspring (probability $1/2$). Similarly, the
probability of changing to a state with $k+1$ females is
$\Omega_{k,k+1}=(N-k)/2N$ and that of remaining with $k$ females is
$\Omega_{k,k}=1/2$. The dynamics is similar to a random walk in the
space of integers $0 \leq k \leq N$ biased towards $k=N/2$: the closer
$k$ is to $0$ the smaller the probability of $k \rightarrow k-1$; the
closer $k$ is to $N$ the smaller the probability of $k \rightarrow
k+1$. This tends to stabilize the population, leading to small
extinction probabilities. We define a generation by $N$ time steps,
which is the average time it takes for the entire population to be
replaced.

The dynamics of $P_t(k)$ is governed by the following equations:
\begin{equation}
\begin{array}{ll}
P_{t+1}(0) &=  \dps{ P_t(0) + \frac{1}{2N} P_t(1)} \\ \\
P_{t+1}(1) &=  \dps{ \frac{N}{2N} P_t(1) + \frac{2}{2N} P_t(2)} \\ \\
P_{t+1}(k) &=  \dps{ \frac{N-k+1}{2N} P_t(k-1) + \frac{N}{2N} P_t(k) +
\frac{k+1}{2N} P_t(k+1), \quad k=2,3, \dots, N-2} \\ \\
P_{t+1}(N\!-\!1) &=  \dps{ \frac{2}{2N} P_t(N-2) + \frac{N}{2N} P_t(N-1)}\\ \\
P_{t+1}(N) &=  \dps{ \frac{1}{2N} P_t(N-1) + P_t(N) }.
\end{array}
\label{master}
\end{equation}

The probabilities $P_t(k)$ define a vector of $N+1$ components
$P_t$, in terms of which the master equation above becomes
\begin{equation}
P_{t+1} = U P_t.
\end{equation}
The evolution matrix $U$ and the transition matrix $\Omega=U^T$ are
tridiagonal. This is a linear system that can be completely solved in
terms of the eigenvalues and eigenvectors of $U$. However, because $U$ is
not symmetric, both right $\vec{a}_r$ and left $\vec{b}_r$ eigenvectors
are needed. Moreover, since $\sum_i U_{ij}= \sum_i \Omega_{j,i}= 1$, $U$
is a {\it stochastic matrix}, having real eigenvalues satisfying
$\lambda_i \leq 1$ and
\begin{equation}
\sum_{r=0}^N  \vec{a}_r . \vec{b}_r^T = \mathbf{1}
\end{equation}
where the superscript $T$ stands for transposition, the lower dot
represents the diadic product and the normalization is set by $\vec{b}_i^T
\cdot \vec{a}_j = \delta_{ij}$. Using this property, the transition
probability between an initial state with $k_0$ females and a state with
$k$ females after the time $t$ can be written as
\begin{equation}
P(k,t;k_0,0) = \sum_{r=0}^N b_{rk_0} a_{rk} \lambda_r^t
\label{problm}
\end{equation}
where $a_{ri}$ is the i-th component of the r-th right eigenvector and
similarly for $b_{rj}$. Equation (\ref{master}) is an example of a Markov
chain, a discrete dynamical system where the transition probability
between any two states depends only on the two states involved and not on
the past history of the system. Markov chains are ubiquitous in genetics
\cite{moran,wat61,cann74,glad78,ewens,gill,jw,aguiar2005,aguiar2007,
wright1943,kimura}, but not so common in population dynamics.

The dynamical system described by Eq.(\ref{master}) is related to the one
dimensional motion of a Brownian particle subjected to an external force
$F(x)$, studied by Smoluchowski and Kac \cite{kac1947}. In this case the
probability $P(x,t)$ of finding the particle at position $x$ at time $t$
satisfies the diffusion equation
\begin{equation}
\frac{\partial P}{\partial t} = D \frac{\partial^2 P}{\partial x^2} -
\frac{1}{f} \frac{\partial}{\partial x} (P F)
\label{smolu}
\end{equation}
where $D$ and $f$ are the diffusion and friction coefficients. In
the limit of large $N$ we may set $\epsilon = 1/N$, $x_k=k \epsilon$ and
the time step to $\delta$ and transform equation (\ref{master}) into a
similar equation,
\begin{equation}
\frac{\partial P}{\partial t} = \frac{\epsilon^2}{4 \delta}
\frac{\partial^2 P}{\partial x^2} + \frac{\epsilon}{\delta} \frac{\partial
}{\partial x} \big[ (x-1/2) P \big].
\label{diff}
\end{equation}
The `diffusion coefficient' is $\epsilon^2/4\delta$ and the `force' is
harmonic, $F(x) = -(x-1/2)$, tending to restore the population towards
$x=1/2$ (or $k=N/2$). The main difference between Eqs.(\ref{master}) and
(\ref{diff}) is that in the former the end points of the diffusion
interval are absorbing states, lending the stationary solution of this
equation unstable for any finite $N$. 

Since each of the $N$ individuals can be either male or female and only
one individual is replaced at each step, the dynamics described by
Eqs.(\ref{master}) can also be mapped into a random walk to nearest
neighbors in a hypercube in $N$ dimensions \cite{diaconis}. This problem,
it turn, is also related to the Ehrenfest model, where $N$ numbered balls
are placed into two boxes and at each time step one ball is chosen at
random and moved to the other box \cite{kac1947}. Starting with all balls
in one of the boxes and letting the system evolve corresponds to the
classic model of a gas confined in one of two chambers and removing
the wall separating the chambers. The state corresponding to all
individuals of the same sex maps into two opposing corners of the
hypercube and to all balls in the same box. 

In this paper tackle the problem of sex ratio fluctuations in population dynamics and
make contact with these classic statistical models. Some of the known analytical
results available in the literature cited above will be connected to the
present calculations below.

%%%%%%%%%%%%%%%%%%%%%%%%%%%%%%%%%%%%%%%%%%%%%%%%%%%%%%%%%
%%%%%%%%%%%%%%%%%%%%%%%%%%%%%%%%%%%%%%%%%%%%%%%%%%%%%%%%%
\section{Absorbing and Transient States}

The eigenvectors corresponding to $\lambda=1$ completely determine the
asymptotic behavior of the system, since the contributions of all the
others die out at long times.

The evolution matrix $U$ is given explicitly by
\begin{equation}
U = \left(
\begin{array}{c|ccccccccc|c}
1 & \frac{1}{2N} & 0 & 0 & 0 & \dots & 0 & 0 & 0 & 0 & 0\\
\hline
0 & \frac{1}{2} & \frac{2}{2N} & 0 & 0 & \dots & 0 & 0 & 0 & 0 & 0\\
0 & \frac{N-1}{2N} & \frac{1}{2} & \frac{3}{2N} & 0 & \dots & 0 & 0 & 0 &
0 & 0\\
\vdots & \vdots & \dots & \dots & \vdots & \dots & \vdots & \vdots &
\vdots & \vdots & \vdots\\
0 & 0 & \dots & \dots & \dots & \dots & 0 &  \frac{3}{2N} & \frac{1}{2}
& \frac{N-1}{2N} & 0 \\
0 & 0 & \dots & \dots & \dots & \dots & 0  & 0  & \frac{2}{2N} &
\frac{1}{2} & 0  \\
\hline
0 & 0 & \dots & \dots & \dots & \dots & 0 & 0 & 0 & \frac{1}{2N} & 1
\end{array} \right).
\label{u}
\end{equation}
There are two eigenvalues $1$, with eigenvectors
\begin{equation}
\begin{array}{cc}
\vec{a}_0 = \left(
  \begin{array}{c}
  1/2 \\ 0 \\ 0 \\ \vdots \\ 0 \\ 1/2
  \end{array} \right) & \qquad \mbox{and} \qquad
\vec{a}_N = \left(
  \begin{array}{c}
  1/2 \\ 0 \\ 0 \\ \vdots \\0 \\ -1/2
  \end{array} \right)
\end{array}
\end{equation}
such that $\vec{a}_0+\vec{a}_N$ corresponds to $P(0)$ and
$\vec{a}_0-\vec{a}_N$ to $P(N)$, both leading to extinction. The
corresponding left eigenvectors are
\begin{equation}
\begin{array}{cc}
\vec{b}_0 = \left(
  \begin{array}{c}
  1 \\ 1 \\ 1 \\ \vdots \\ 1 \\ 1
  \end{array} \right) & \qquad \mbox{and} \qquad
\vec{b}_N = \left(
  \begin{array}{c}
  1 \\ b_{N,1} \\ b_{N,2} \\ \vdots \\b_{N,N-1} \\ -1
  \end{array} \right).
\end{array}
\end{equation}
The vector $\vec{b}_N$ does not have a simple form for finite $N$ but the
coefficients $b_{N,k}$ go to zero for large $N$ and $k=1,2 \dots, N-1$.
Notice that this choice of vectors agrees with $\vec{b}_i^T \cdot
\vec{a}_j = \delta_{ij}$ in this sub-space. Using these vectors,
equation (\ref{problm}) can be re-written, for $k_0 \neq 0, N$, as
\begin{equation}
P(k,t;k_0,0) = \frac{1}{2} \delta_{k,0} + \frac{1}{2} \delta_{k,N} +
\sum_{r=1}^{N-1} b_{rk_0} a_{rk} \lambda_r^t \label{problm2}
\end{equation}
making explicit that extinction is the only asymptotic possibility,
independent of the initial state $k_0$.

The other eigenvalues and eigenvectors can be calculated from
the non-trivial $(N-1)\times(N-1)$ part of $U$, delimited by the lines
in equation (\ref{u}),
\begin{equation}
V  = \frac{1}{2}+ \frac{1}{2N} W
\end{equation}
where
\begin{equation}
W = \left(
\begin{array}{ccccccccc}
0 & 2 & 0 & 0 & \dots & 0 & 0 & 0 & 0\\
N-1 & 0 & 3 & 0 & \dots & 0 & 0 & 0 & 0\\
0 & N-2 & 0 & 4 & \dots & 0 & 0 & 0 & 0\\
\vdots & \dots & \dots & \vdots & \dots & \vdots & \vdots & \vdots &
\vdots\\
0 & \dots & \dots & \dots & \dots & 4 &  0 & N-2 & 0 \\
0 & \dots & \dots & \dots & \dots & 0 &  3 & 0 & N-1 \\
0 & \dots & \dots & \dots & \dots & 0  & 0  & 2 & 0
\end{array} \right).
\end{equation}
If $\lambda_i$ are the eigenvalues of $U$, and $\mu_i$ the eigenvalues of
$W$, then
\begin{equation}
 \lambda_i = \frac{1}{2}\left(1+ \frac{\mu_i}{N} \right).
\end{equation}

%%%%%%%%%%%%%%%%%%%%%%%%%%%%%%%%%%%%%%%%%%%%%%%%%%%%%%%%%
%%%%%%%%%%%%%%%%%%%%%%%%%%%%%%%%%%%%%%%%%%%%%%%%%%%%%%%%%
\section{Eigenvalues and eigenvectors for large $N$}

Although $W$ has a very simple structure, its eigenvalues cannot be
calculated analytically for arbitrary $N$. It can be checked that the
resulting polynomial for the eigenvalues $\mu$ gets more and more
complicated as $N$ increases. However, in the limit $N \rightarrow
\infty$, we find
\begin{equation}
\mu_i = N - 2(i-1) \qquad i=1,2, \dots.
\end{equation}
The largest eigenvalue, $\mu_1= N$, yields a new asymptotically
stable state, with $\lambda_1=1$ and eigenvector
\begin{equation}
a_{1k} = c_1 \exp{\left[-2N\left( \frac{k}{N}-\frac{1}{2}
\right)^2\right]},
\label{a1k}
\end{equation}
corresponding to a symmetric distribution of males and females centered
at $k=N/2$. In the limit of large $N$ we may set $\epsilon = 1/N$, $x_k=k
\epsilon$ and obtain 
\begin{equation}
a(x) = \sqrt{\frac{2}{\pi \epsilon}}
\exp{\left[-\frac{2}{\epsilon}\left(x-\frac{1}{2}\right)^2\right]},
\end{equation}
which is the stationary solution of equation (\ref{diff}).
For finite $N$, however, $\lambda_1 < 1$ and the population
inevitably goes extinct towards $k=0$ or $k=N$. Therefore, it is important
to estimate how $\lambda_1$ tends to 1 as $N$ goes to infinity, a
calculation presented in the next section. This provides an estimate
of the time to extinction due to fluctuations in the sex determination.

All other values of $\mu_i$ lead to $\lambda_i < \lambda_1$ and do not
contribute to the asymptotic  state of the population. The eigenvectors
are given by
\begin{equation}
a_{ik} = c_i \exp{\left[-2N\left( \frac{k}{N}-\frac{1}{2}
\right)^2\right]} H_{i-1}(\sqrt{2N}(k/N-1/2)),
\end{equation}
where $H_m(x)$ are the Hermite polynomials and $c_i$ are normalization
constants. The appearance of Hermite polynomials in the solution can be
traced to the diffusion equation (\ref{diff}) and its interpretation via 
Smoluchowski's Brownian motion. These results are demonstrated in the
appendix \ref{appev}.

%%%%%%%%%%%%%%%%%%%%%%%%%%%%%%%%%%%%%%%%%%%%%%%%%%%%%%%%%
%%%%%%%%%%%%%%%%%%%%%%%%%%%%%%%%%%%%%%%%%%%%%%%%%%%%%%%%%
\section{Largest eigenvalue of $W$ for finite $N$}

Writing $\mu_1 = N - \alpha$, the eigenvalue equation
$\det{[W-\mu_1 \mathbf{1}]}=0$ becomes
\begin{equation}
\det{[C+\alpha \mathbf{1}]} = \det{[C]}
\det{[\mathbf{1}+\alpha C^{-1}]} = 0,
\end{equation}
where $C=W-N\mathbf{1}$. Using
\begin{equation}
\det{[1+C^{-1}\alpha]} = 1 + \alpha Tr[C^{-1}] + {\cal O}(\alpha^2)
\label{detapp}
\end{equation}
we obtain $\alpha=-\left\{Tr[C^{-1}]\right\}^{-1}$,
\begin{equation}
\mu_1 = N + \left\{Tr[C^{-1}]\right\}^{-1}
\label{corrmu1}
\end{equation}
and
\begin{equation}
\lambda_1 = 1 + \frac{1}{2N} \left\{Tr[C^{-1}]\right\}^{-1}.
\label{corrlamb1}
\end{equation}

\begin{figure}
\includegraphics[scale=0.6]{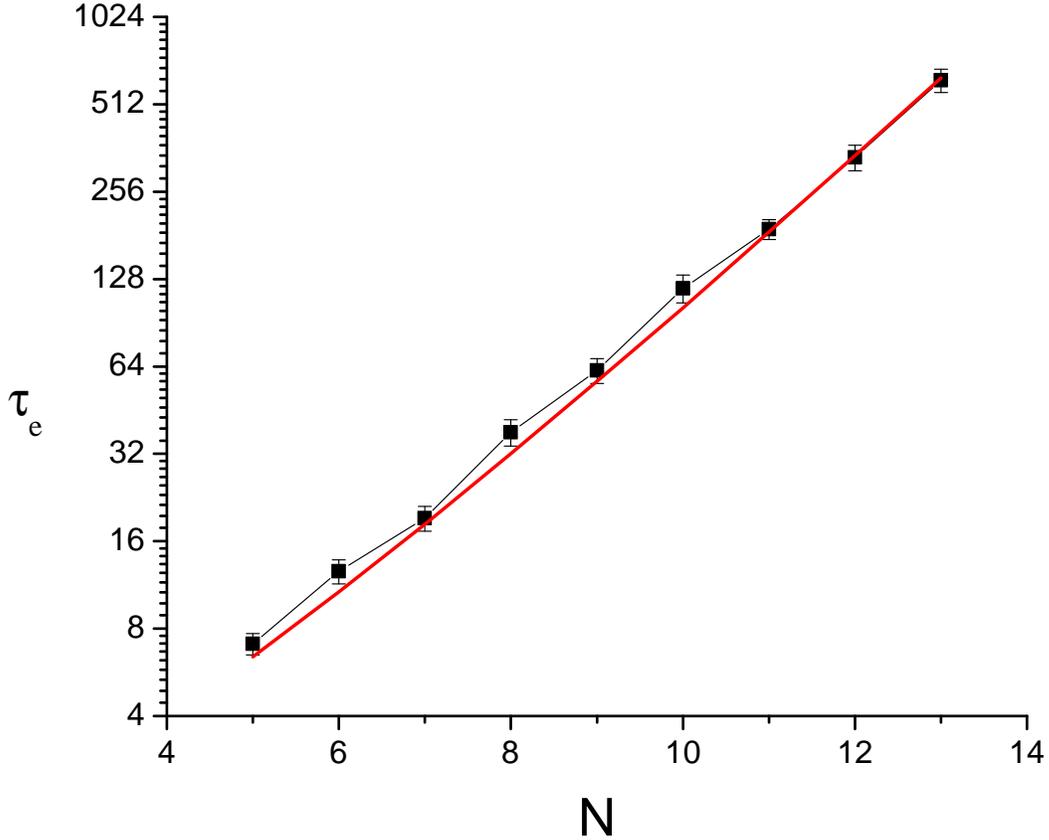}
\caption{(Color online) Time to extinction as a function of $N$. The
black points show the average and mean square deviation obtained with 100
replicates of simulations. The red line shows the theoretical result, Eq.
(\ref{tau}).}
\label{fig1}
\end{figure}

We show in Appendix \ref{appc} that
\begin{equation}
Tr[C^{-1}] = \sum_{k=1}^{N-1} \frac{f_{k-1} f_{N-k-1}}{f_{N-1}}
\end{equation}
where
\begin{equation}
f_k = -N f_{k-1} - k(N-k+1)f_{k-2}
\end{equation}
with $f_0=1$ and $f_1=-N$. It turns out that, for large $N$,
\begin{equation}
\frac{f_{k-1} f_{N-k-1}}{f_{N-1}} = -\frac{1}{N} B(N,k)[1+{\cal O}(1/N)]
\label{nm1}
\end{equation}
where $B(N,k)$ is the binomial coefficient. In this approximation the sum
can be easily performed and the result is
\begin{equation}
 \lambda_1 = 1 - 2^{-(N+1)} \approx 1 - 2^{-N},
 \label{lambda1}
\end{equation}
which is very accurate even for small $N$. Setting $\lambda_1 =
e^{-1/\tau_e}$ we obtain the time to extinction as $2^N$ time steps. In
terms of number of generations
\begin{equation}
 \tau_e =  2^{N} / N.
 \label{tau}
\end{equation}
The last approximation in Eq. (\ref{lambda1}), where we multiply by a
factor 2, is justified because Eq.(\ref{nm1}) is accurate only up to the
scaling behavior of order $1/N$. The factor 2 is obtained by fitting the
numerical simulations (see figure \ref{fig1}). The time to extinction can
be related to Ehrenfest model. In this problem $N$ balls are placed in
two boxes and a random ball is moved from its box to the other at each
time step. Starting from an arbitrary state, the time it takes for finding
all the balls in one box is of order $2^N$ \cite{kac1947}. This is also
the Poincar\'e recurrence time, which is the time it takes to return to
the state with all balls in the same box, having started there. 

Considering the meta-stable probability distribution $P(k) = 2^{-N}
B(N,K)$ corresponding to $\lambda_1$, one could ask how long it takes
for an arbitrary initial state to reach $P(k)$. This is given by the
next eigenvalue, $\lambda_2=1-1/N$, with the associated relaxation time
of a single generation:
\begin{equation}
 \tau_r =  1.
 \label{taur}
\end{equation}
This is the second important time scale of the problem, much shorter
than $\tau_e$. This result is analogous to that obtained in
\cite{diaconis} for the time taken by a particle to reach the stationary
distribution on the hypercube under a nearest neighbor random walk, which
is of order $N\log{N}$.

As a simple example of these time scales, a population with $N=20$
individuals starting with $k=10$ females fluctuates according to the
binomial distribution after 1 generation, or $20$ time steps, but goes
extinct only after about $\tau_e = 2^{20}/20 \approx $ fifty thousand
generations (or a million time steps). Figure \ref{fig1} shows a
comparison between numerical calculations and the theoretical prediction
of $\tau_e$, showing good agreement even for small values of $N$.

%%%%%%%%%%%%%%%%%%%%%%%%%%%%%%%%%%%%%%%%%%%%%%%%%%%%%%%%%
%%%%%%%%%%%%%%%%%%%%%%%%%%%%%%%%%%%%%%%%%%%%%%%%%%%%%%%%%
\section{Robustness against changes in offspring sex ratio}

In some species the birth of one of the sexes is favored over the other. A
well studied example is the infection caused by the bacteria Wolbachia,
that kills males in some arthropod species \cite{hurst99}. Sex
determination can also be influenced by temperature in several species of
reptiles, like the snow skink lizard \cite{pen2010}. Changes in the sex
ratio of offspring not only shifts the distribution of males and females
in the population but also affects the time to extinction due to random
fluctuations. In this section we discuss the time to extinction assuming
that the probability of a female offspring is $p=1/2 + s$ and that of a
male is $1-p=1/2-s$.

As we demonstrate below, the inclusion of bias in the sex ratio at birth
complicates the dynamics and only approximate solutions for the time to
extinction can be derived. However, the problem can be easily solved for
the Wright-Fisher model, where generations are non-overlapping and
constructed from the previous one by independent random choice of males
and females. The Wright-Fisher model describes, for instance,
annual plants, where the entire population dies in the winter and its
replaced anew in the spring. The Moran model, on the other hand, is
appropriate for perennial plants. The probability of $k$ females is given
by 
\begin{equation}
 (1/2+s)^k (1/2-s)^{N-k} B(N,k) 
 \label{wfp}
\end{equation}
and the probability of extinction is, therefore $(1/2+s)^N + (1/2-s)^N$,
which reduces to $2^{-(N-1)}$ for $s=0$. The time to extinction is the
inverse of this probability and is already given in terms of number of
generations: $2^{(N-1)} = (2^N/N) \times (N/2)$. Although this
can be taken as a first estimate for the time to extinction in the Moran
model, it overestimates it by a factor $(N/2)$ (see Eq.(\ref{tau})). The
factor N comes from updating the entire population at once. Numerical
simulations indicate that the factor 1/2, valid for s=0 only, becomes
smaller for more extremes values of s. 

In the case of the Moral model, the transition matrix elements are
generalized to
\begin{equation}
 \begin{array}{l}
\Omega_{k,k-1}=\frac{k}{2N}(1-2s)\\
\\
\Omega_{k,k}=\frac{k}{2N}(1+2s)+\frac{N-k}{2N}(1-2s)=\frac{1}{2
N}(N+2s(2k-N))\\
\\
\Omega_{k,k+1}=\frac{N-k}{2N}(1+2s)\\
\end{array}
\end{equation}
and a master equation similar to equation (\ref{master}) can be written.
In the extreme cases $s=\pm 1/2$ all the eigenvalues of the evolution
matrix $U$ can be calculated analytically and the largest non-unit
eigenvalue is $\lambda_1=1-1/N$, so that the time to extinction in
approximately $N$, i.e., 1 generation.

In the limit of large $N$, the master equation can also be  written as a
diffusion equation

\begin{equation}
\frac{\partial P}{\partial t} = \frac{\partial}{\partial x}
\left[ D(x,s) \frac{\partial P}{\partial x}  \right] - 
\frac{\epsilon}{\delta} \frac{\partial }{\partial x} \big[ F(x,s) P
\big]
\label{diffs}
\end{equation}

\noindent with a 'space-dependent' diffusion coefficient $D(x,s) =
(\epsilon^2/4\delta)(1+2s-4sx)$ and $F(x,s)=-(x-1/2-s-\epsilon s)$. The
stationary solution is

\begin{equation}
P(x) = A \exp{\left\{\frac{4sx+[1-4(1+\epsilon)s^2] \ln(1+2s-4sx)}
{4\epsilon s^2} \right\}} ,
\label{stats}
\end{equation}
where $A$ is a normalization constant. Notice the symmetry of these
equations with respect to the change $s \rightarrow -s$ and $x
\rightarrow 1-x$.

Differential equations for the continuous limit of the coefficients $a_k$
and $b_k$ can also be obtained and the corresponding solutions are
$\mu_i = N - 2(i-1)$ with
\begin{equation}
b_i(x)= c_i  H_{i-1}\left(\sqrt{\frac{2N}{1-4s^2}}(x-1/2-s)\right)
\label{bis}
\end{equation}
and
\begin{equation}
a_i(x)= d_i   \exp{\left\{-\frac{2N(x-1/2-s)^2}{1-4s^2}\right\}} b_i(x),
\label{ais}
\end{equation}
where $H_i(x)$ are the Hermite polynomials. Although the solution for
$a_1(x)$ looks rather different from (\ref{stats}), they are very similar
for large $N$. Equation (\ref{ais}) for $i=1$, corresponding to
$\mu_1=N$, is just a Gaussian centered at $x=1/2+s$ with variance
$1-4s^2$, as it should be.

\begin{figure}
\includegraphics[scale=0.6]{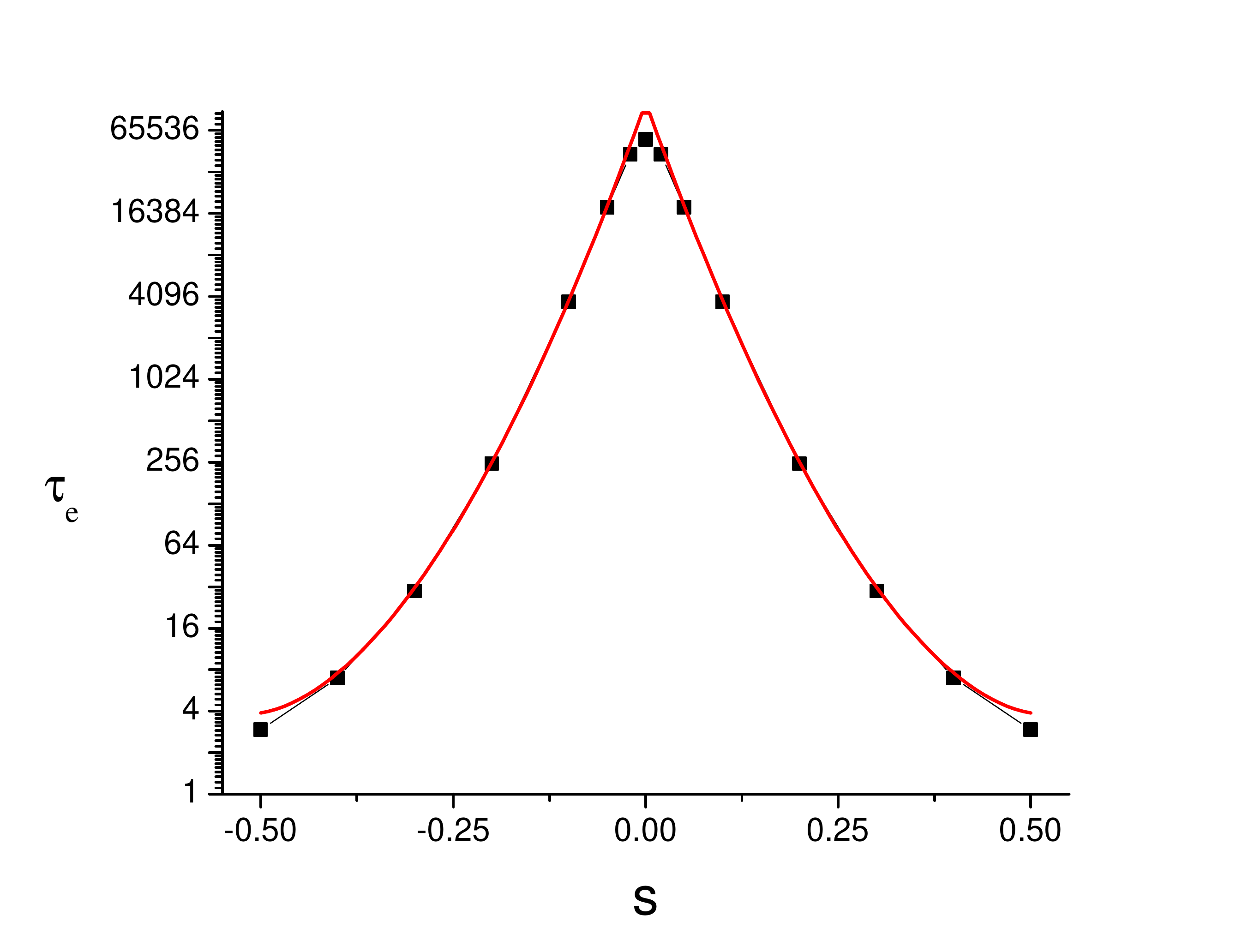}
\caption{(Color online) Time to extinction as a function of $s$ for a
population with $N=20$ starting with 10 males and 10 females. The line
with square symbols shows the result of simulations obtained from 10,000
realizations for each value of $s$. The thick red line shows the
approximation $\tau_2 = a 2^{N_{eff}}/N_{eff}$ for $a=2$.}
\label{fig2}
\end{figure}

The correction to the $\mu_1$ is still given by equation (\ref{corrmu1})
but now with
\begin{equation}
Tr[C^{-1}] = \sum_{k=1}^{N-1} \frac{f_{k-1}(-s) f_{N-k-1}(s)}{f_{N-1}(s)}
\equiv \sum_{k=1}^{N-1} d(N,k,s)
\label{dnks}
\end{equation}
where
\begin{equation}
f_k(s) = -[N+2s(N-2k)] f_{k-1}(s) - k(N-k+1)(1-4s^2)f_{k-2}(s)
\end{equation}
with $f_0=1$ and $f_1=-N+2s(N-2)$. Explicit evaluation of the
coefficients $d(N,k,s)$, however, is much more complicated and an analytic
expression is not available. Numerical simulations show that, for large
$N$, we may approximate
\begin{equation}
d(N,k,s) = -\frac{A(N,s)}{N}
\exp{\left\{-\frac{2N(k/N-1/2-s)^2}{1-4s^2}\right\}}
\end{equation}
where
\begin{equation}
A(N,s) = 2^{N (1-1.4|s|)^2}
\end{equation}
was obtained by fitting the amplitude of the coefficients. Further
approximating the sum over $k$ in equation (\ref{dnks}) by an integral
we obtain
\begin{equation}
Tr[C^{-1}] = -A(N,s) \sqrt{\frac{\pi(1-4s^2)}{2N}}
\end{equation}
so that the time to extinction becomes
\begin{equation}
\tau_e =  2 \sqrt{\frac{\pi(1-4s^2)}{2N}}
2^{N (1-1.4|s|)^2} \sim 2^{N_{eff}}
\end{equation}
with $N_{eff}=N(1-1.4|s|)^2$. Figure 2 shows the time to extinction as
a function of $s$ for $N=20$. The solid line shows the result of
simulations and the red line a fit with $\tau_2 = a 2^{N_{eff}}/N_{eff}$.
The goodness of the fit suggests that the expression is correct and that
the main effect of $s$ is to change the population size to an effective
value.

Our calculations have shown that producing males and females offspring at
the same proportion is a good strategy even for very small populations.
If, however, the birth of one of the sexes is favored, not only the
balance between males and females is altered but the time to extinction
might decrease dramatically. For small deviations, however, the
exponential character of the extinction time remains.

%%%%%%%%%%%%%%%%%%%%%%%%%%%%%%%%%%%%%%%%%%%%%%%%%%%%%%%%%%%
%%%%%%%%%%%%%%%%%%%%%%%%%%%%%%%%%%%%%%%%%%%%%%%%%%%%%%%%%%%
\begin{acknowledgments}
It is a pleasure to thank Ayana Martins and Carolina Reigada for
pointing out the relevance of considering bias in the offspring sex ratio.
This work was partly supported by FAPESP (EC, DMS and MAMA) and CNPq
(MAMA).
\end{acknowledgments}

\begin{appendix}
%%%%%%%%%%%%%%%%%%%%%%%%%%%%%%%%%%%%%%%%%%%%%%%%%%%%%%%%%%%
%%%%%%%%%%%%%%%%%%%%%%%%%%%%%%%%%%%%%%%%%%%%%%%%%%%%%%%%%%%
\section{Eigenvalues and eigenvectors of $W$ for large $N$}
\label{appev}

Setting $a_{i0}=a_{iN}=b_{i0} = b_{iN}=0$ we obtain the following
recurrence relations for the i-th right and left eigenvectors of $W$:
\begin{equation}
\label{recurrencia}
(k+1)a_{i k+1}+(N-k+1)a_{i k-1}-\mu_i a_{i k} =0
\end{equation}
and
\begin{equation}
(N - k)b_{i k+1} + k b_{i k-1} - \mu_i b_{i k}=0.
\label{bequ}
\end{equation}
It can be checked that these components are related by
\begin{equation}
a_{i k} = 2^{-N}Bi(N,k) b_{i k},
\label{aequ}
\end{equation}
where $Bi(N,K)$ is the binomial coefficient.

Dividing equation (\ref{bequ}) by $N$ and defining the function
$\tilde{b}_i(x)$ such that $\tilde{b}_i(x_k) = b_{i k}$ leads to
\begin{equation}
(1 - x)\tilde{b}_i (x + \epsilon) + x \tilde{b}_i(x - \epsilon) -
\frac{\mu_i}{N}\tilde{b}_i(x) =0,
\label{bdiffeq}
\end{equation}
where $\epsilon = 1/N$.

For  $N\gg 1$, we can approximate
$\tilde{b}_i(x\pm\epsilon)=\tilde{b}_i(x) \pm\epsilon
\tilde{b}'_i(x)+\frac{\epsilon^2}{2}\tilde{b}''_i(x)$. Neglecting terms
of $O(\epsilon^3)$, equation (\ref{bdiffeq}) is turned into a
differential equations for $b(x)$:
\begin{equation}
\tilde{b}^{\prime\prime}_i\frac{\epsilon^2}{2} +
\tilde{b}^{\prime}_i\epsilon(1 - 2x) + \tilde{b}_i\left(1 -
\frac{\mu_i}{N}\right)=0.
\label{diferenc}
\end{equation}

Taking the limit $N\rightarrow \infty$ ($\epsilon \rightarrow 0$) we see
that $\mu_1=N$ is an eigenvalue. The corresponding left and right
eigenvectors are given by

\begin{equation}
b_{1 k} = 1, \ \ \  a_{1 k} = 2^{-N} \frac{N!}{k!(N - k)!} \textrm{ for }
k = 1 \cdots N - 1.
\label{bmu1}
\end{equation}

To obtain the remaining eigenvalues and eigenvectors for large $N$ we
define $t\equiv\sqrt{2N}(x-1/2)$ and the function $g(t)\equiv
\tilde{b}(\frac{t+1/2}{\sqrt{2 N}})$. Accordingly, equation
(\ref{diferenc}) becomes
\begin{equation}
g_i''-2 t g_i' + 2 d_i g_i =0
\label{eqDifG}
\end{equation}
where $d_i\equiv (N-\mu_i)/2$. Equation (\ref{eqDifG}) is satisfied by
the Hermite polynomials if $d_i$ were integers. To see that this is
indeed the case, note that the boundary conditions are
$\tilde{b}(0)=\tilde{b}(1)=0$, where the argument is
$x=t/\sqrt{2N}+1/2$. Therefore, the corresponding boundary
conditions for $g(t)$ are $g(\pm\sqrt{N/2})=0$ or, for
$N\rightarrow\infty$, $g(\pm\infty)=0$. In order to prevent $g(t)$ from
diverging we must set $d_i=0,1,2,\ldots$. The procedure is similar to the
quantization of the harmonic oscillator in quantum mechanics. In this
limit the eigenvalues and eigenfunctions are given by
\begin{equation}
\mu_i=N-2 (i-1)
\end{equation}
and
\begin{equation}
g_i(t)=c_i H_{i-1}(t)
\label{solution_R}
\end{equation}
or
\begin{equation}
b_{ik}= c_i  H_{i-1}(\sqrt{2N}(k/N-1/2))
\label{a_i_k}
\end{equation}
where $i=1,2,3,\ldots$ and the $c_i$ are normalization constants.

For the right-eigenvectors we can approximate the binomial by the normal
distribution $Bi(p,N) \approx N(\mu,\sigma)$ by taking $ p =1/2$, $\mu =
N p$ and $\sigma = Np(1-p)$. We obtain
\begin{equation}
 a_{i k} = c^{\prime}_i \exp[2N(k/N - 1/2)^2] H_{i-1}(\sqrt{2N}(k/N-1/2)).
\end{equation}
%

%%%%%%%%%%%%%%%%%%%%%%%%%%%%%%%%%%%%%%%%%%%%%%%%%%%%%%%%%%%
%%%%%%%%%%%%%%%%%%%%%%%%%%%%%%%%%%%%%%%%%%%%%%%%%%%%%%%%%%%
\section{The trace of $C^{-1}$}
\label{appc}

The matrix $C$ is given by
\begin{equation}
C = \left(
\begin{array}{ccccccccc}
-N & 2 & 0 & 0 & \dots & 0 & 0 & 0 & 0\\
N-1 & -N & 3 & 0 & \dots & 0 & 0 & 0 & 0\\
0 & N-2 & -N & 4 & \dots & 0 & 0 & 0 & 0\\
\vdots & \dots & \dots & \vdots & \dots & \vdots & \vdots & \vdots &
\vdots\\
0 & \dots & \dots & \dots & \dots & 4 &  -N & N-2 & 0 \\
0 & \dots & \dots & \dots & \dots & 0 &  3 & -N & N-1 \\
0 & \dots & \dots & \dots & \dots & 0  & 0  & 2 & -N
\end{array} \right).
\end{equation}
and we only need the diagonal elements of $C^{-1}$, which can be obtained
by Laplace's formula:
\begin{equation}
 [C^{-1}]_{kk} = \frac{Cofactor(k,k)}{Det[C]}.
\end{equation}
The Cofactor(i,j) is the determinant of the auxiliary matrix obtained
by removing the j-th line and i-th row of $C$, multiplied by
$(-1)^{i+j}$.

The removal of the k-th line and row of $C$ divides the remaining
matrix into two decoupled blocks. In order to deal with these blocks we
recursively define the matrix $C_{N-k}$ to be the matrix $C_{N-k+1}$
with the first line and first row removed. This definition holds for
$k=2,\ldots,N-1$, with the initial condition $C_{N-1}\equiv C$. Calling
$f_{N-k} = Det[C_{N-k}]$ it is easy to see that
\begin{equation}
 Cofactor(k,k) = f_{k-1}f_{N-k-1}.
\label{cofac}
\end{equation}
Applying the Laplace rule to the determinant $f_k$ it can be checked
that it satisfies the recurrence relation
\begin{equation}
f_k = -N f_{k-1} - k(N-k+1)f_{k-2}
\end{equation}
with $f_0\equiv 1$ and $f_1 \equiv -N$.

In order to calculate (\ref{cofac}) it is useful to define
$g_k=(-1)^{k+1} f_{k-1}/k!$. In terms of $g_k$ the recurrence relation
becomes
\begin{equation}
(k+1)g_{k+1}  -N g_k + (N-k+1)g_{k-1}=0
\label{recg}
\end{equation}
with $g_0\equiv 0$ and $g_1 \equiv 1$. We obtain
\begin{equation}
 [C^{-1}]_{kk} = \frac{f_{k-1} f_{N-k-1}}{f_{N-1}} = -\frac{g_k ~ g_{N-k}}{g_N} \frac{1}{B(N,k)}
\label{cofacg}
\end{equation}
where $B(N,k)$ is the binomial coefficient. Since (\ref{recg}) is the
relation satisfied by $B(N,k)$ itself, it is reasonable to assume
that
\begin{equation}
g_k = \frac{1}{N} B(N,k)
\label{gsol}
\end{equation}
where the factor $1/N$ guarantees the initial condition $g_1=1$. This,
however, is only an approximation, since it gives $g_0=1/N$ and not
zero. However, for large $N$ it suffices for obtaining the first order
correction to the eigenvalue. Replacing (\ref{gsol}) into
(\ref{cofacg}) we obtain
\begin{equation}
 [C^{-1}]_{kk} = - \frac{B(N,k)}{N}.
\label{cofacf}
\end{equation}

\end{appendix}

\bibliographystyle{unsrt}

\end{document}